\documentclass[onecolumn,showpacs,preprintnumbers,amsmath,amssymb]{revtex4}
\usepackage{graphicx}
\usepackage{dcolumn}
\usepackage{bm}
\makeatletter
\newcommand\figcaption{\def\@captype{figure}\caption}
\makeatother

\newcommand{\ez}{{\boldsymbol e}_{z}}

\begin{document}

\title{Formation of robust and completely tunable resonant photonic band gaps}

\author{Shiyang  Liu}
\author{Junjie Du, Zhifang Lin}
\affiliation{Surface Physics Laboratory, Department of Physics,
Fudan University, Shanghai 200433, China}

\author{Ruixin Wu}
\affiliation{Department of Electronic Sciences and Engineering,
Nanjing University, Nanjing 210093, China}

\author{S. T. Chui}
\affiliation{Bartol Research Institute, University of Delaware,
Newark, Delaware 19716, USA}

\date{\today}

\begin{abstract}
We identify different types of the
photonic band gaps (PBGs) of two dimensional magnetic photonic
crystals (MPCs) consisting of arrays of magnetic cylinders and study
the different tunability (by an external static magnetic field) of
these PBGs.  One type of the band gaps comes from infinitely
degenerate flat bands and is closely related to those in the study
of plasmonics. In addition, such PBGs are magnetically tunable and
robust against position disorder. We calcualte the transmission of
the PBG's and found excellent agreement with the results of the
photonic band structure calculation. Positional disorder of the
lattice structure affects the different types of PBGs differently.
\end{abstract}

 \maketitle

\newpage
The
unique characteristic of photonic crystals (PCs)
is the existence of photonic band gaps (PBGs), the frequency
ranges over which electromagnetic (EM) modes are forbidden.
%
For applications in optical devices, it is important to obtain
tunable PCs
so that the photonic band structures or the PBGs can be modulated
\cite{tiaoc252} and, preferably, robust.
%
Magnetic photonic crystals (MPCs) have attracted interest\cite{sigprb56}
because of the potential tunability of the PBGs by an external
static magnetic field (ESMF) and the
fast switching time of magnetic systems.
Very recently Chen \textit{et al}
found experimentally that the ESMF
can vary the position of the PBGs \cite{xujpd40,chejpcm19}, but by
only a small amount. This raises doubt on the practical
tunability of the MPCs.

In this paper we demonstrate for realistic experimental parameters
that even though not all PBGs are easily tunable, there are PBGs
that can easily be controlled by ESMFs. We found three kinds of
PBGs: (1) from the Bragg scattering; (2) from the Mie scattering
resonance of magnetic cylinders, and (3) from the spin wave
resonance. The responses of these to ESMFs differ considerably. Both
the 2nd and the 3rd types of bands are very sensitive to the ESMF
and are magnetically tunable whereas the first type is not.
We calcualte the transmission of the PBG's under different ESMFs and
found excellent agreement with the results from the photonic band
structure. The transmission as a function of lattice position
disorder is also studied. Positional disorder only affects the Bragg
type of PBGs significantly but not the other two, implying the
robust tunable PBGs. We now describe our results in detail.

For bands with flat dispersion the plane wave
expansion method does not work well,
in our work
the multiple scattering method is employed.  The two
dimensional (2D) MPCs are composed of hexagonal lattices of soft
Mg-Mn\cite{chejpcm19} ferrite rods
in nonmagnetic plexiglas, with lattice constant $a$.
The axis of the cylinder and the ESMF ${\bf H}_{ext}=H_{ext}\ez$ are along
the $z$ direction, the wave vector is in the xy plane.
For ferrite rods, the relative permeability tensor of
the ferrite and its inverse are
\begin{equation}
\widehat{\mu}=\mu_s \left(\begin{array}{ccc}
\mu_r&-i\mu_\kappa&0\\
i\mu_\kappa&\mu_r&0\\
0&0&1
\end{array}\right),
\mu_s\widehat{\mu}^{-1}=\left(\begin{array}{ccc}
\mu'_r&-i\mu'_\kappa&0\\
i\mu'_\kappa&\mu'_r&0\\
0&0&1
\end{array}\right),
\label{mu}
\end{equation}
where $\mu_s=1$, 
$\mu_s\mu_r=1+\omega_m(\omega_0-i\alpha\omega)/[(\omega_0-i\alpha\omega)^2-\omega^2]$, 
$\mu_s\mu_\kappa=\omega_m\omega/[(\omega_0-i\alpha\omega)^2-\omega^2]$,
$\omega_m=\gamma 4\pi M_s$ with $M_s$ the saturation magnetization, $\omega_0=\gamma
H_0$ is the ferromagnetic resonance frequency with $\gamma$
the gyromagnetic ratio, $H_0$ is the sum of the external field and the shape anisotropy field, $\alpha$ is the damping coefficient of the ferrite. 
$\mu'_r=\mu_r/(\mu_r^2-\mu_\kappa^2)$, 
$\mu'_\kappa=-\mu_\kappa/(\mu_r^2-\mu_\kappa^2).
$

The time dependence of the field is assumed to be $e^{-i\omega t}$
and suppressed. In 2D EM systems both the transverse
electric (TE) mode and the transverse magnetic (TM) mode are the
eigen-modes.
Only the TM mode is tunable by the ESMF and will be considered here. To
illustrate the physics, we consider a concrete example where the
lattice constant is taken as $a=8mm$, the radius of the ferrite
cylinder is taken as $r_s=\frac{1}{4}a=2mm$. The relative
permittivity and permeability of the nonmagnetic background medium
(plexiglas) are $\epsilon_m=2.6$ and $\mu_m=1$. For the ferrite
cylinder $\epsilon_s=12.3,$ the saturation magnetization $4\pi
M_s=1700\text{G}$. We have neglected the absorption of the ferrite
cylinder, the imaginary part of $\widehat{\mu}$ of the ferrite in
the microwave region is very weak.

\begin{figure}[t]
\includegraphics[width=0.40\textwidth,angle=0]{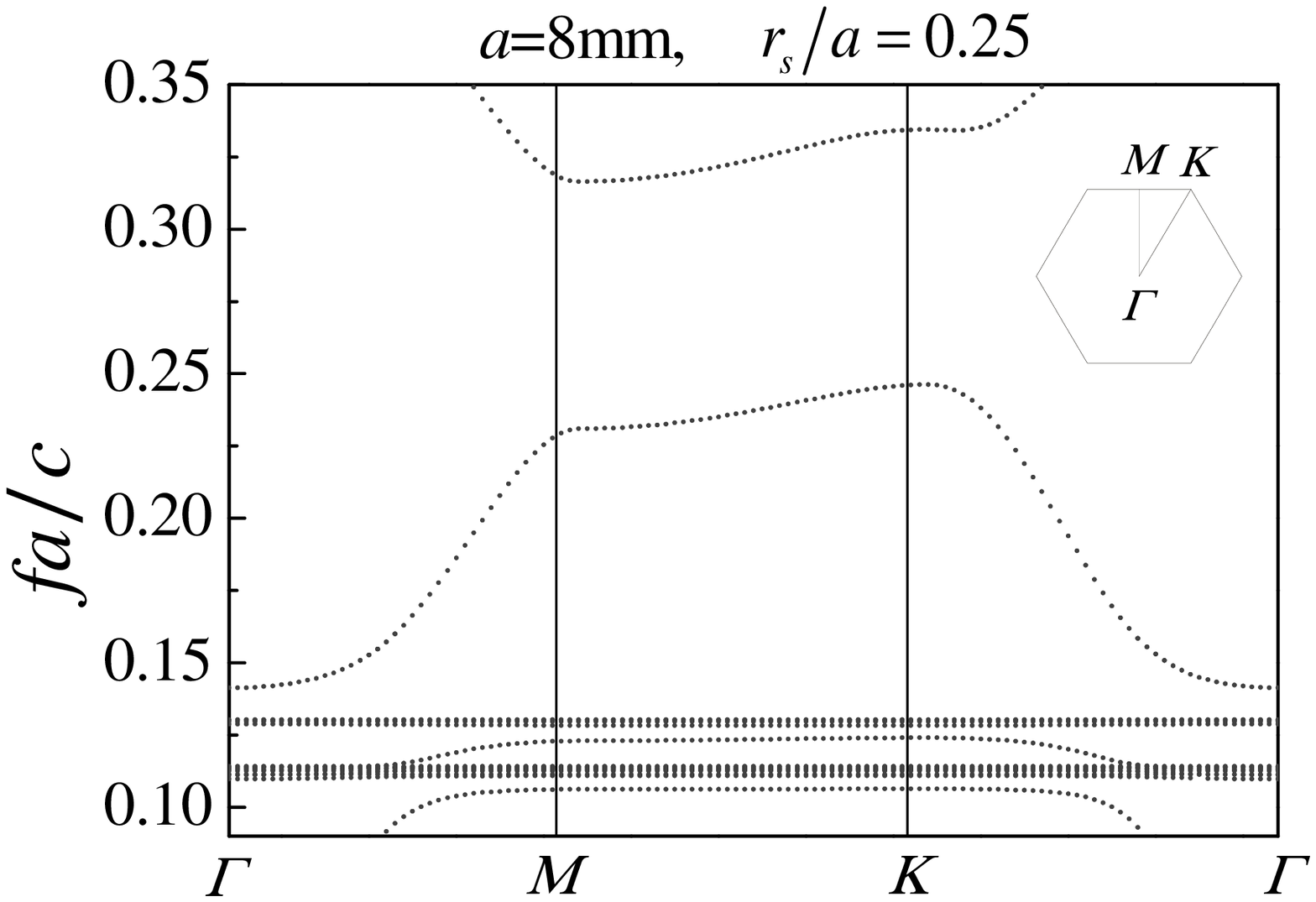}
\caption{\label{a8r2h900t} The photonic band structure of a two
dimensional MPC. $a=8mm$ and $r=2mm$, $H_0=900\text{Oe}$. The
reduced Brillouin zone (RBZ) is illustrated in the upper right hand
corner. The coordinates of the high symmetry points appearing in the
RBZ are ${\it\Gamma}=\frac{2\pi}{a}(0,0)$, ${\it
M}=\frac{2\pi}{a}(0,\frac{1}{\sqrt{3}})$ and
$K=\frac{2\pi}{a}(\frac{1}{3},\frac{1}{\sqrt{3}})$.}
\end{figure}

\begin{figure}
\includegraphics[width=0.45\textwidth,height=3cm,angle=0]{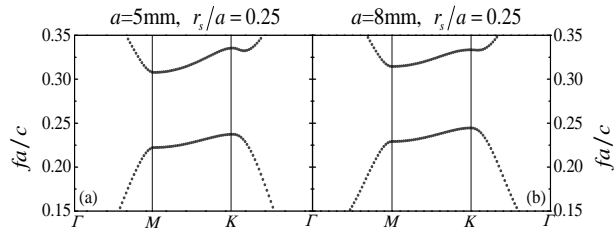}
\caption{\label{a5a8r125r2h700b} The photonic band structure near the
top PBG  for (a) $a=5mm$ and $r_s=1.25mm$; (b) $a=8mm$ and $r_s=2mm$.
The other parameters are the same as those in Fig.\,\ref{a8r2h900t}.}
\end{figure}

In Fig\,\ref{a8r2h900t} we show the photonic band structure
with a ESMF $H_0=900\text{Oe}$.
The frequency is in units of $c/a$ with $c$ the speed of light
in vacuum.
The PBGs are located near
$0.108c/a$, $0.125c/a$, $0.135c/a$ and $0.275c/a$, respectively.
In conventional dielectric PCs, the PBGs could be understood
as a result of Bragg scattering.
%
The location of a PBG due to
Bragg scattering
scales with the inverse lattice constant.
In Fig.\,\ref{a5a8r125r2h700b} we show
the photonic band structure near the top PBG for
lattice constants (a) $a=5mm$, $r=\frac{1}{4}a=1.25mm$
and (b) $a=8mm$, $r=\frac{1}{4}a=2mm$; the
other parameters are the same as those in Fig.\,\ref{a8r2h900t}.
When expressed in units of c/a,
the PBGs appeared in almost the same position, in addition the photonic
band structures are nearly the same on both sides of the
PBGs. This provides convincing evidence that scale invariance
is satisfied by this PBG. We conclude that this PBG
is dominated by Bragg scattering.

The photonic band structures near the top PBG
for different ESMFs of $H_0=700\text{Oe}$ and  $H_0=900\text{Oe}$
are nearly the same as each
other.
This explains why there is only a
tiny change in the positions of the PBGs when different ESMFs are
applied in the experiments of Chen \textit{et al}
\cite{xujpd40,chejpcm19}, which is performed in this frequency range.
%
We next turn our attention to
the other PBGs.

The "Mie" scattering resonance of cylinders for {\bf all}
angular momenta $n>0$ occurs when
$\mu_{eff}=\mu_r+\mu_\kappa=\mu_r'+\mu_\kappa'=-1$
($\epsilon=-1$) for TM (TE) modes in the long wavelength limit.
We found analytically\cite{unpublished} that this corresponds
to infinite number of degenerate
flat bands {\bf in 2D} when the cylinder radius is small.
For scattering from spheres in 3D, the resonance conditions are different
for different angular momenta and the 3D bands are nondegenerate.
This kind of flat bands for TE modes have been discussed in the
context of plasmonics\cite{plasmonics} where the focus is on the fact that
$\epsilon=-1$ ($\epsilon=-2$ in 3D for s wave)
corresponds to the resonance condition
of surface plasmons.  $\mu_{eff}=-1$ at a frequency
$f_s=\frac{1}{2\pi}[\omega_0+\frac{1}{2}\omega_m].$
For our parameters this corresponds to frequencies of
$0.1158c/a$ and $0.1308c/a$ for ESMFs of
$H_0=700\text{Oe}$ and $H_0=900 \text{Oe}$.

\begin{figure}
\includegraphics[width=0.40\textwidth,height=3cm,angle=0]{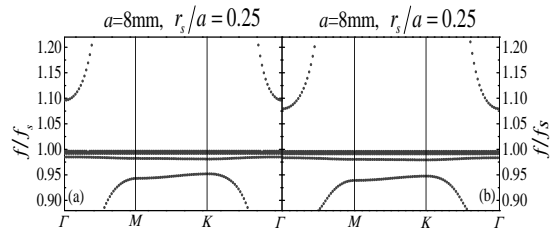}
\caption{\label{a8r2h700900mr} Photonic band structure near the
second and the third PBGs with the frequency divided by $f_s$. The
parameters are the same as those in Fig.\,\ref{a8r2h900t} except
that (a) $H_0=700\text{Oe}$; (b) $H_0=900\text{Oe}$.}
\end{figure}

In Fig.\,\ref{a8r2h700900mr} we plotted the photonic band structures
near the second and the third PBG
with the eigenfrequencies divided by $f_s$ for different ESMFs.
The scaling of the frequencies of
these bands with $f_s$ is obvious.
ESMFs change  $f_s$ through $\omega_0$ and hence the positions of
the second and the third PBGs.
The position of the PBG can be switched to a new position without
a PBG originally.
A propagating wave can become totally reflected and vice versa
by changing the ESMF. This phenomenon can be useful in the design of
optical devices.
%
The PBG near these flat bands are not associated with the Bragg
scattering dominating the conventional dielectric PBG.
We perofrmed calculations of
the photonic band structures near these flat bands for
different lattice constants a=5mm and 8mm and found that
the flat band between the second and
the third PBGs are located at the same position and does
{\bf not} scale with $1/a$. 
Our understanding suggests that if disorder is introduced in the periodicity the
first PBG will be affected seriously while the second and the third
will persist.

There exists another flat band in Fig.\,\ref{a8r2h900t} lying
between the third and the fourth PBG. The position of this band
can also be varied by an ESMF.
A "spin wave" resonance occurs at a
frequency $f_m=\frac{1}{2\pi}\sqrt{\omega_0(\omega_0+\omega_m)}$
when $\mu_r=\mu_r'=0$ and the wavevector inside the cylinder,
which is proportional to $1/\sqrt{\mu_r'}$, becomes infinite.
For $H_0=900\text{Oe}$ and $4\pi M_s=1700\text{G}$
$f_m=0.114c/a$ corresponding to the bottom of the flat band
in Fig\,\ref{a8r2h900t}. This fourth PBG
can be manipulated by an ESMF, which maybe just as useful as the second and the
third PBGs. 

\begin{figure}
\includegraphics[width=0.4\textwidth,angle=0]{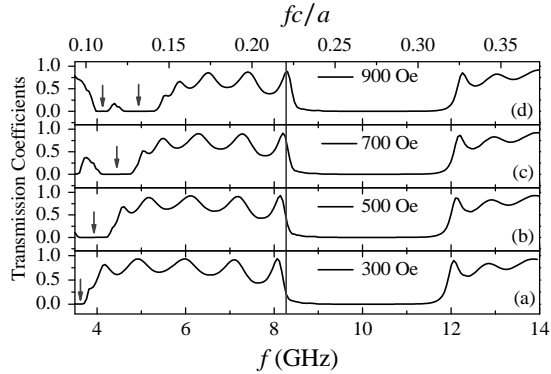}
\caption{\label{a8r2h300-900Transmission} Transmission coefficients
of the sample exerted with different ESMFs. The lattice constant is
$8mm$ and the radius of the ferrite cylinder is
$r_s=\frac{1}{4}a=2mm$. (a) $H_0=300\text{Oe}$; (b)
$H_0=500\text{Oe}$; (c) $H_0=700\text{Oe}$; (d) $H_0=900\text{Oe}$.}
\end{figure}

For comparison with the experimental result it is necessary to
perform the simulation of transmission coefficients. This can be
expressed as the forward scattering amplitude of the
MPC\cite{liprb58}. For this calculation, the small imaginary part of
the realistic susceptibilities are included:
$\epsilon_b=2.6+i0.005,$
$\epsilon_s=12.3+i0.0006$,
$\alpha=0.007.$
%
In Fig\,\ref{a8r2h300-900Transmission}
we show the transmission for different ESMFs for a $7$-layer
structure normal to the $x$ direction with each layer consisting of
$13$ ferrite cylinder along $y$ direction
$a=8mm,$ $r=\frac{1}{4}a=2mm$.
For
convenience, the frequency are given in units of GHz
(bottom)
and $c/a$ (top).
The locations of the second and
the third PBGs are labeled with arrows in the figure.
With the
increase of the ESMF these PBG rise from the low frequency up
to high frequency significantly.
However, for the first PBG only a tiny change occurs with the
variation of ESMFs from $300\text{Oe}$ to $900\text{Oe}$. This can
be seen from the vertical line denoting the low frequency edge of
the first PBG.
The positions of the PBGs of the transmission coefficients can be
compared with those in the photonic band structures. For example, in
Fig.\,\ref{a8r2h300-900Transmission} (d) the middle gap frequency of
the first, the second and the third PBGs are at about $0.275c/a$,
$0.135c/a$ and $0.125c/a$, respectively, the same as those shown in
the photonic band structures in Fig.\,\ref{a8r2h900t}.
In addition, the width of the PBGs in
transmission coefficients and photonic band structures are almost
the same. In Fig.\,\ref{a8r2h300-900Transmission} we find some
oscillation of the transmission away from the PBG. This is due to
the finite size of the sample, as is pointed by Stefanou \textit{et
al.} \cite{stecpc113}. The same kind of results are obtained for
$a=5mm$, $r=a/4.$

\begin{figure}[t]
\includegraphics[width=0.48\textwidth,height=3cm,angle=0]{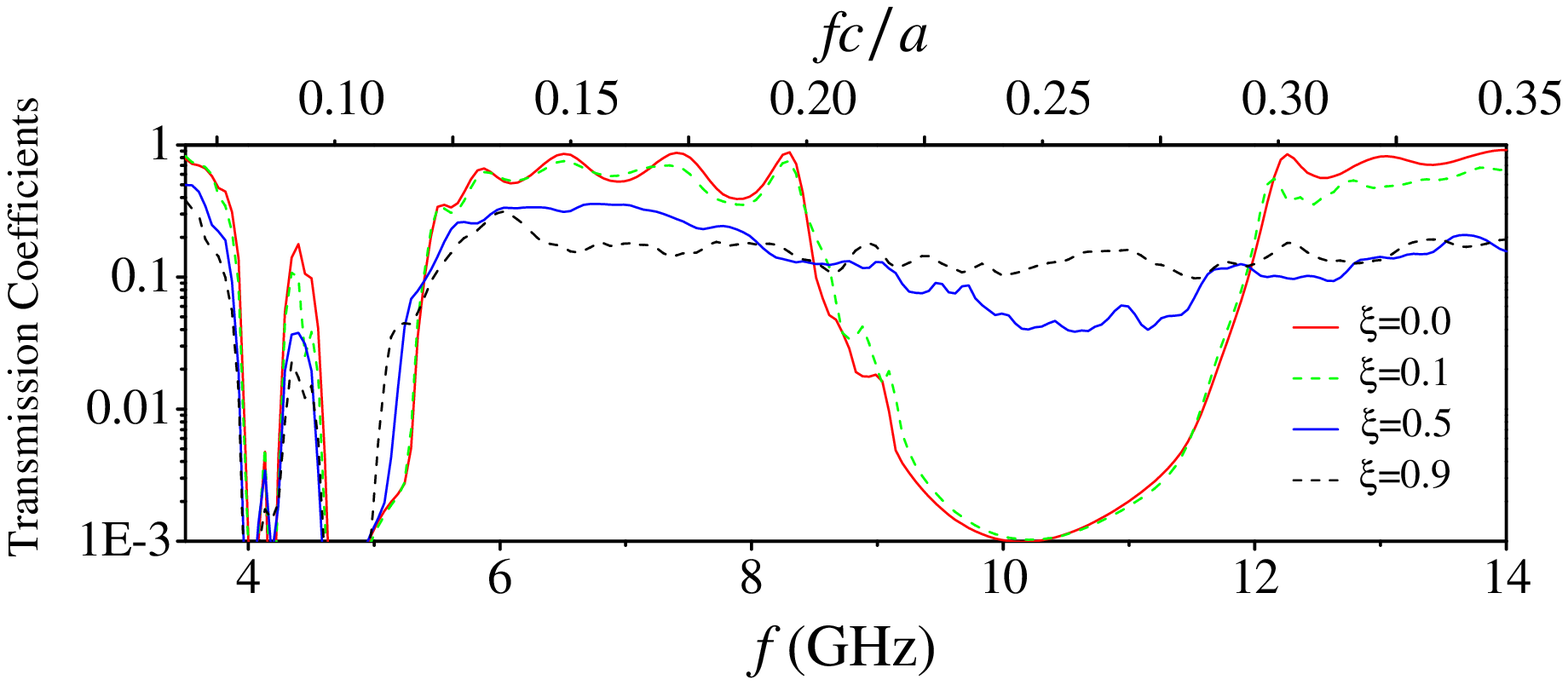}
\caption{\label{a8r2h900ran0-0.1-0.5-0.9transmission} Transmission
coefficients of the sample with the introduction of disorder. The
ESMF exerted is $H_0=900\text{Oe}$. The lattice constant is $8mm$
and the radius of the ferrite cylinder is $r_s=\frac{1}{4}a=2mm$.}
\end{figure}

In our approach, randomness of magnitude $\xi$ can
be introduced for the positions of the ferrite cylinder.
%
Along $x$ and $y$ directions the maximum variations of the position
of the ferrite cylinder are $d_{x0}=\frac{\sqrt{3}}{2}a-2r$ and
$d_{y0}=a-2r$ respectively, so that the displacement of each
cylinder is $d_i=d_{i0}\xi\nu_i$ with $i=x\ \text{or}\ y$ and $\nu_i$ are
random numbers between 0 and 1. In
Fig.\,\ref{a8r2h900ran0-0.1-0.5-0.9transmission} the results of the
transmission coefficients with different randomness $\xi$ are
presented.
There exists
no obvious change in the transmission coefficent and the PBGs when
the disorder is very weak.
The oscillation of the transmission coefficient off the PBG becomes
irregular. With the increase of the randomness ($\xi=0.5$) the first
PBG disappears due to the breakdown of the Bragg scattering; the
"magnetic" PBG remain relatively unchanged.
The larger the randomness, the less the
transmission coefficient becomes.

In summary, we have performed photonic band structure
calculations and simulations of the transmission coefficients with
and without randomness.
Four PBGs 
are found. 
The top PBG originates
from the Bragg scattering
and is sensitive to position disorder,
the second and the third PBGs, from the
magnetic Mie scattering resonance
of the cylinder and the fourth PBG, from the
"spin wave" resonance. Even though the Bragg PBG is insensitive to
an external field, the magnetic PBGs
are easily tunable, are robust aganist position
disorder and thus are suitable for
application in designing optical devices.
%

This work was supported by CNKBRSF, NNSFC through grant No. 10474014
and PCSIRT. RXW is supported by NNSFC through grant No. 60771013.
STC is partly supported in part by the US DOE.

\newpage

\begin{thebibliography}{99}
\bibitem{tiaoc252} H. P. Tian, and J. Zi, Opt. Commun. {\bf 252}, 321 (2005).
\bibitem{sigprb56} M. M. Sigalas, C. M. Soukoulis, R. Biswas, and K. M. Ho, Phys. Rev. B {\bf 56}, 959 (1997).
E. D. V. Nagesh, V. Subramanian, V. Sivasubramanian, and V. R. K. Murthy, Physica B {\bf 382}, 45 (2006).
I. L. Lyubchanskii, N. N. Dadoenkova, M. I. Lyubchanskii, E. A. Shapovalov, and Th. Rasing, J. Phys. D {\bf 36}, R277 (2003).
Z. Lin, and S. T. Chui, Phys. Rev. E {\bf 69}, 056614 (2004).
S. Liu, and Z. Lin, Phys. Rev. E {\bf 73}, 066609 (2006). Z. Lin,
and S. T. Chui, Opt. Lett. {\bf 32} 2288 (2007). S. T. Chui, and Z.
Lin, J. Phys.: Condens. Matter {\bf 19} 406233 (2007).
\bibitem{xujpd40} J. Xu, R. X. Wu, P. Chen, and Y. Shi, J. Phys. D: Appl. Phys. {\bf 40}, 960 (2007).
\bibitem{chejpcm19} P. Chen, R. X. Wu, J. Xu, A. M. Jiang, and X. Y. Ji, J. Phys.: Condens. Matter {\bf 19}, 106205 (2007).
\bibitem{plasmonics}
A.V. Zayats, I.I. Smolyaninov, A.A. Maradudin, "Nano-optics of surface
plasmon polaritons," Phys. Rep., vol. 408, pp. 131-314 (2005).
\bibitem{unpublished} S. T. Chui, Z. F. Lin and L. Zhou, unpublished
\bibitem{liprb58} L. M. Li, and Z. Q. Zhang, Phys. Rev. B {\bf 58}, 9587 (1998).
\bibitem{stecpc113} N. Stefanou, V. Yannopapas, and A. Modinos, Compu. Phys. Commun. {\bf 113}, 49 (1998).
\end{thebibliography}
\end{document}